\documentclass[12pt]{article}
\usepackage{graphicx}
\usepackage{caption}
\usepackage{subcaption}

\newcommand\pubnumber{DPF2015-152}
\newcommand\pubdate{\today}

\def\napoli{Department of Physics\\
University of Texas at Arlington, Arlington, TX76019, USA}
\def\support{\footnote{Work supported by the Office of Science, 
          U.S. Department of Energy, under award DE-SC0011686.}}

\def\Title#1{\begin{center} {\Large #1 } \end{center}}
\def\Author#1{\begin{center}{ \sc #1} \end{center}}
\def\Address#1{\begin{center}{ \it #1} \end{center}}

\newcommand\pubblock{\rightline{\begin{tabular}{l} \pubnumber\\
         \pubdate  \end{tabular}}}
\newenvironment{Abstract}{\begin{quotation}  }{\end{quotation}}
\newenvironment{Presented}{\begin{quotation} \begin{center} 
             PRESENTED AT\end{center}\bigskip 
      \begin{center}\begin{large}}{\end{large}\end{center} \end{quotation}}

\begin{document}
\begin{titlepage}
\pubblock

\vfill
\Title{The SiD Detector for the International Linear Collider}
\vfill
\Author{ Andrew P. White\support
(for the SiD Consortium)}
\Address{\napoli}
\vfill
\begin{Abstract}
The SiD Detector is one of two validated detector designs for the future International Linear Collider. 
SiD features a compact, cost-constrained design for precision Higgs couplings determination, and other measurements, and sensitivity 
to a wide range of possible new phenomena. A robust silicon vertex and tracking system, combined with a 5 Tesla 
central solenoidal field, provides excellent momentum resolution. The highly granular calorimeter system is 
optimized for Particle Flow application to achieve very good jet energy resolution over a wide range of energies. 
Details of the proposed implementation of the SiD subsystems, as driven by the physics requirements, will be given. 
The shared interaction point, push-pull mechanism, will be described, together with the estimated timeline for construction.
\end{Abstract}
\vfill
\begin{Presented}
DPF 2015\\
The Meeting of the American Physical Society\\
Division of Particles and Fields\\
Ann Arbor, Michigan, August 4--8, 2015\\
\end{Presented}
\vfill
\end{titlepage}
\def\thefootnote{\fnsymbol{footnote}}
\setcounter{footnote}{0}

\section{Introduction}

\begin{figure}[htb]
\centering
\includegraphics[height=2.0in]{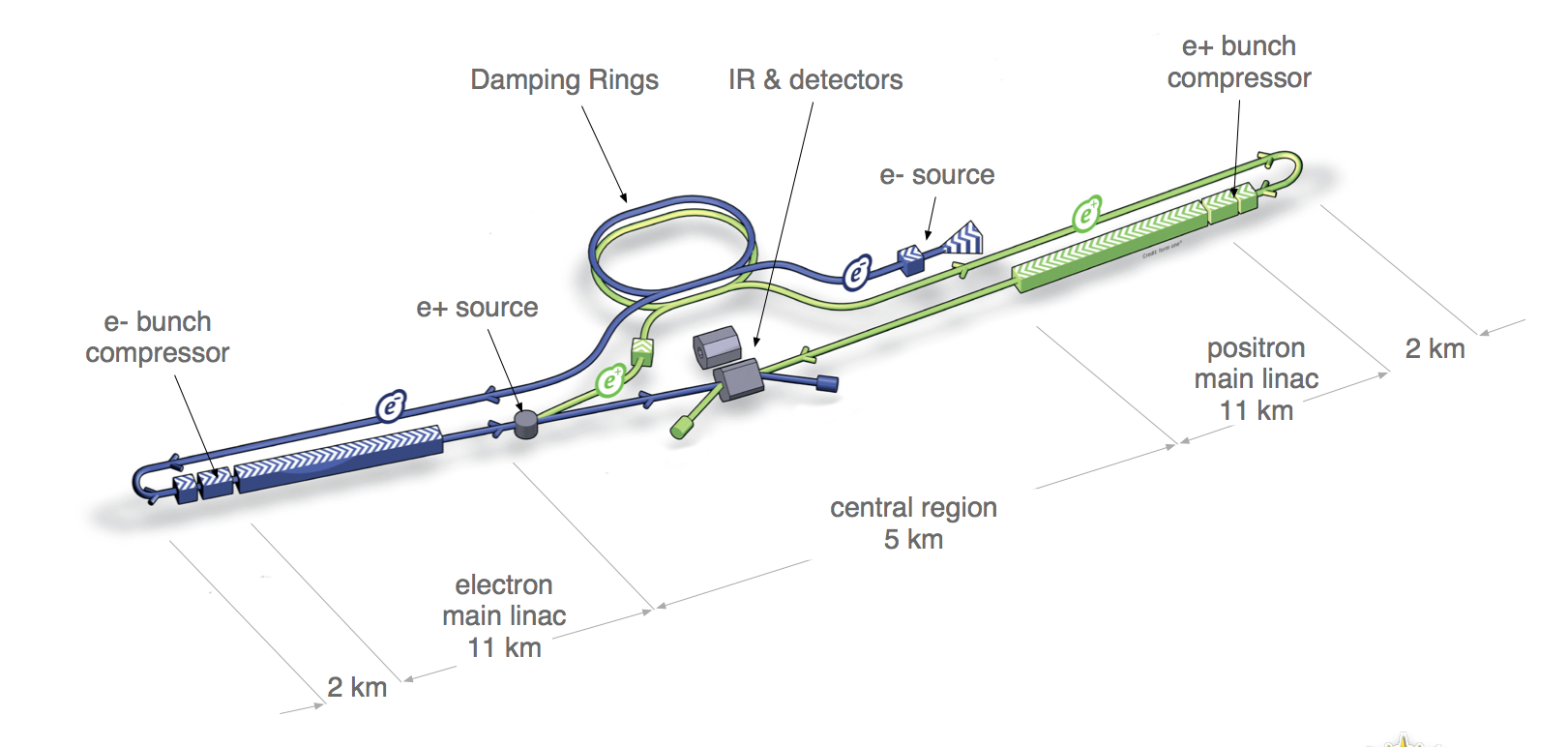}
\caption{The ILC accelerator showing the location of the detectors.}
\label{fig:ILC_Acc}
\end{figure}

The SiD Detector is one of two detector concepts developed for the future 
International Linear Collider, to be located in Japan at the proposed Kitakami
site. The ILC accelerator is shown in Fig.~\ref{fig:ILC_Acc}.

The SiD detector is designed for a comprehensive program
of e+e- physics ranging from percent-level measurement of Higgs boson couplings to precision top quark
studies, to searches for new physics such as supersymmetry. The detector requirements 
to address this physics program are given in Table \ref{ilcdbd:tab:sub-performance}. Many of the Higgs 
processes require excellent jet energy performance which derives from the design and performance 
of the tracking and calorimeter systems together with an efficient particle-flow algorithm.
The Higgs recoil mass measurement requires excellent charged particle momentum resolution,
while Higgs branching fraction measurements for heavy flavors rely on a precision vertex
detector. Finally, searches for new phenomena require a hermetic
design.

\begin{table}[h]
\caption{Detector performance needed for key ILC physics measurements. \label{ilcdbd:tab:sub-performance} }

\begin{center}

\begin{tabular}{lllll}



\hline
\multicolumn{1}{l}{Physics} & \multicolumn{1}{l}{Measured} & \multicolumn{1}{l}{Critical} & \multicolumn{1}{l}{Physical} & \multicolumn{1}{l}{Required}  \\
\multicolumn{1}{l}{Process} & \multicolumn{1}{l}{Quantity} & \multicolumn{1}{l}{System} & \multicolumn{1}{l}{Magnitude} & \multicolumn{1}{l}{Performance}  \\

\hline
\scriptsize Zhh & \scriptsize Triple Higgs coupling  & \scriptsize Tracker & \scriptsize Jet  Energy & \\
\scriptsize $Zh \rightarrow q\bar{q}b\bar{b}$ & \scriptsize Higgs mass & \scriptsize \ \ and & \scriptsize Resolution & \\ 
\scriptsize $Zh \rightarrow ZWW^*$  & \scriptsize $B(h \rightarrow WW^*$) & \scriptsize Calorimeter & \scriptsize $\Delta E/E$ &  \scriptsize 3\% to 4\%\\
\scriptsize $\nu\overline{\nu} W^+W^-$ & \scriptsize $\sigma(e^+e^- \rightarrow\nu\overline{\nu} W^+W^-)$ & & & \\
\hline 
\scriptsize $Zh \rightarrow \ell^+ \ell^- X$  &  \scriptsize Higgs recoil mass  &  \scriptsize $\mu$ detector & \scriptsize Charged \ \ \ \  particle& \\
\scriptsize $ \mu^+ \mu^- (\gamma)$  & \scriptsize Luminosity weighted E$_{\rm {cm}}$ & \scriptsize Tracker & \scriptsize Momentum Resolution & \scriptsize $5 \times 10^{-5} (GeV/c)^{-1}$\\ 
\scriptsize $Zh + h \nu \overline{\nu} \rightarrow \mu^+ \mu^- X$  & \scriptsize BR($h \rightarrow \mu^+ \mu^-$) &  & \scriptsize $\Delta p_t / p^2_t$ &  \\
\hline
\scriptsize $Zh, h \rightarrow b \bar{b}, c\bar{c},  b \bar{b}, gg $  \scriptsize & \scriptsize Higgs branching fractions  &  \scriptsize Vertex & \scriptsize Impact & \scriptsize $5 \mu m \oplus $ \\
  & \scriptsize b-quark charge asymmetry &   & \scriptsize parameter & \scriptsize $10\mu m /p\rm{(GeV/c)}sin^{3/2}\theta$\\ 
\hline
 &   &  \scriptsize Tracker & \scriptsize Momentum Resolution &  \\
 \scriptsize SUSY, eg. $\tilde{\mu}$ decay & \scriptsize $\tilde{\mu}$ mass  &  \scriptsize Calorimeter & \scriptsize Hermeticity &  \\
 &  &  \scriptsize  $\mu$ detector & &  \\
\hline

\end{tabular}
\end{center}
\end{table}


\section{The SiD Detector Concept Design}

\begin{figure}[htb]
\centering
\includegraphics[height=2.5in]{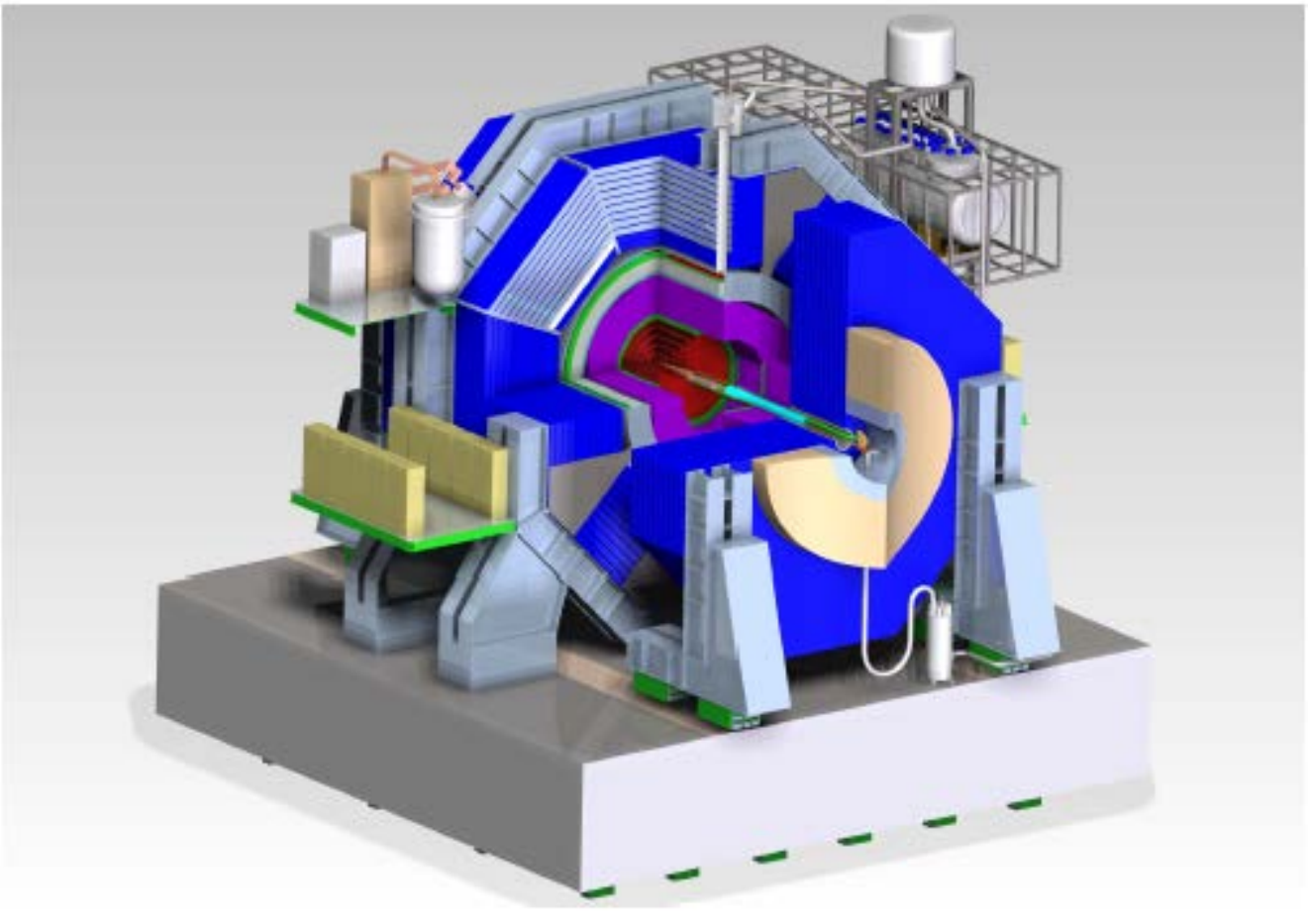}
\caption{The SiD Detector Concept.}
\label{fig:SiD_det}
\end{figure}

SiD is a compact, cost-constrained detector designed to make precision measurements 
and be sensitive to a wide range of new phenomena. The compact design is achieved
by the use of a high precision silicon vertexing and tracking system in combination with a
5 Tesla solenoidal magnetic field. The vertexing and tracking system offers excellent
charged particle momentum resolution and is live for single bunch crossings. 
The calorimetry is optimized for jet energy resolution, based on a particle flow (PFA) approach, 
with tracking calorimeters, compact showers in the electromagnetic section, and highly segmented, 
longitudinally and transversely, electromagnetic and hadronic systems. 
The iron flux return and muon identifier is a component of SiD self-shielding.
The complete detector system is designed for push-pull operation.
The main elements of the SiD detector are shown in Fig.~\ref{fig:SiD_det}.


\section{SiD Vertexing and Tracking}
\begin{figure}
\centering
\begin{subfigure}{.5\textwidth}
\centering
\includegraphics[width=.7\linewidth]{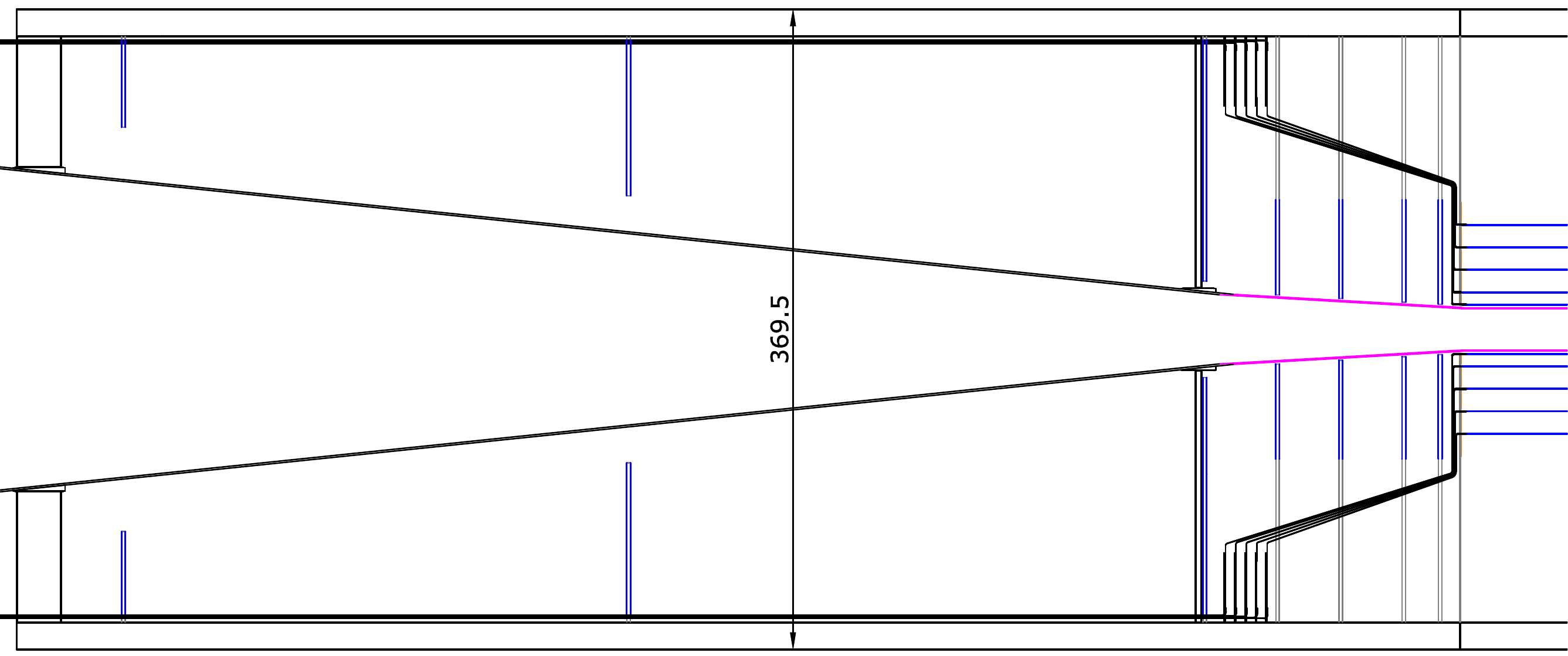}
\caption{Half view of the SiD Vertex Detector.}
\label{fig:vtx_half_barrel}
\end{subfigure}%
\begin{subfigure}{.5\textwidth}
\centering
\includegraphics[width=.7\linewidth]{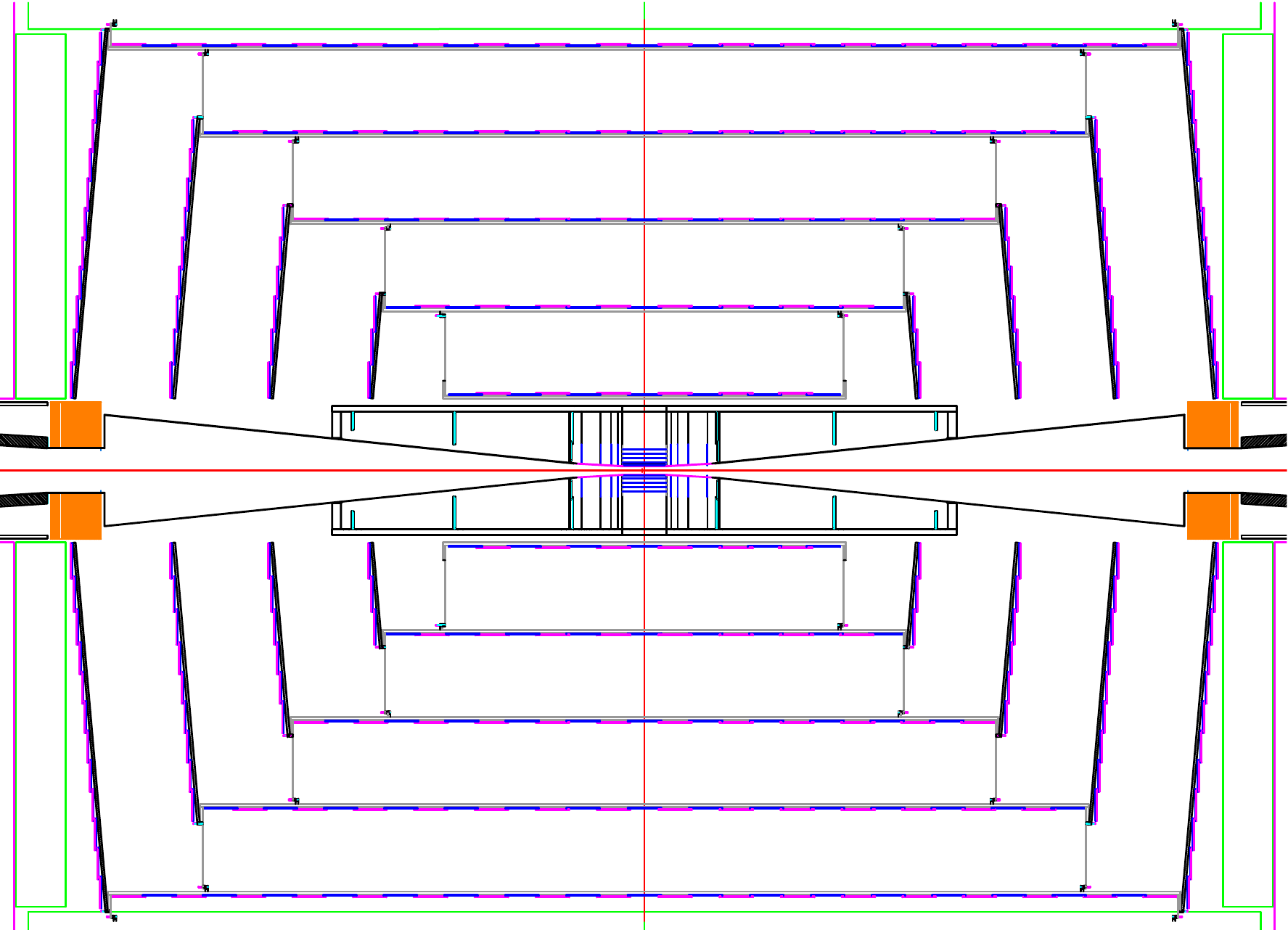}
\caption{SiD Main tracking system.}
\label{fig:trk_overall}
\end{subfigure}
\caption{SiD tracking system - outer tracker (right) and vertex detector (left).}
\label{fig:vtx_trk}
\end{figure}

The SiD vertex detector, Fig.~\ref{fig:vtx_trk} (left), is an all-silicon system
consisting of five barrel layers, four disk layers, and three additional small 
pixel disks in the forward region. The carbon fiber support structure is connected to 
the beam tube in four places.

Various technologies are being considered for the vertex detector, ranging from 
standard silicon diode pixels, through monolithic active pixels (as in the Chronopix~\cite{Chrono} design ),
to vertically integrated 3-dimensional structures~\cite{VIP_chip}. 
Power management for the vertex detector will take advantage of the ILC beam time
structure to use pulsed power, and will use DC-DC conversion to avoid the needed
for high mass cables that would compromise the otherwise excellent low material
profile.

The main tracker is also an all-silicon system with the cylindrical barrel layers 
closed at the ends by conical, annular disks, as shown in Fig.~\ref{fig:vtx_trk} (right).

The main tracker layers are instrumented with silicon microstrip tiles read out
via the KPiX~\cite{KPiX} ASIC which features a four-deep pipeline and single bunch
time stamping with readout occurring between bunch trains. Power pulsing and DC-DC conversion
allows the use of gas cooling and cable mass reduction. 

\begin{figure}[htb]
\centering
\includegraphics[height=2.5in]{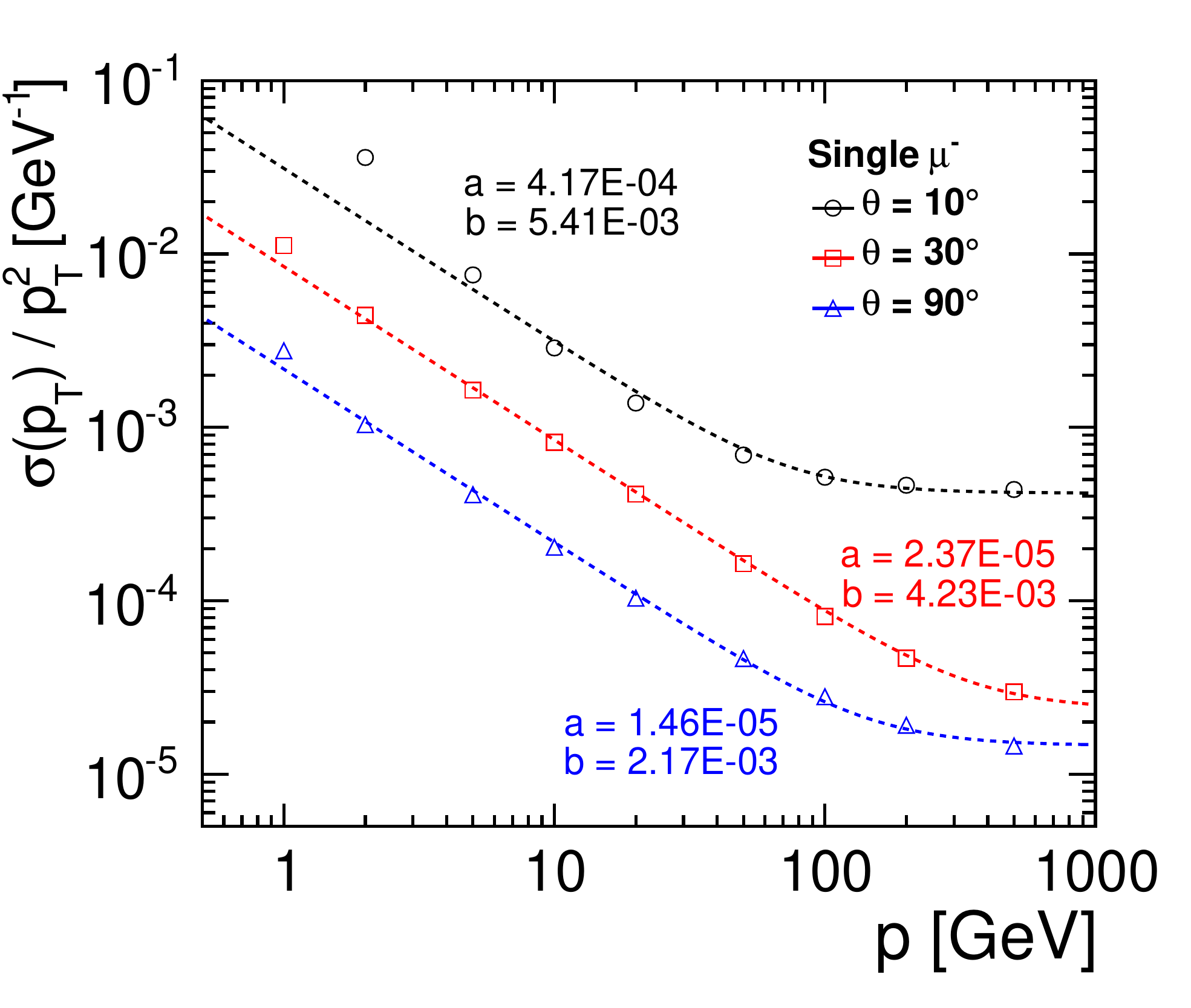}
\caption{Normalized transverse momentum resolution for single muons in SiD.}
\label{fig:mom_resolution}
\end{figure}

Overall for the tracking system, better than 20\% of a radiation length is achieved
over the angular range to within 10 degrees of the beam direction. A tracking efficiency
of at least 95\% is achieved over a similar angular range for charged particles with 
momenta above 1~GeV. The transverse momentum resolution for single muons is shown in
Fig.~\ref{fig:mom_resolution}, which, for central tracks, exceeds the requirement in
Table \ref{ilcdbd:tab:sub-performance}.

\section{Calorimetry}
\begin{figure}
\centering
\begin{subfigure}{.45\textwidth}
\centering
\includegraphics[width=.7\linewidth]{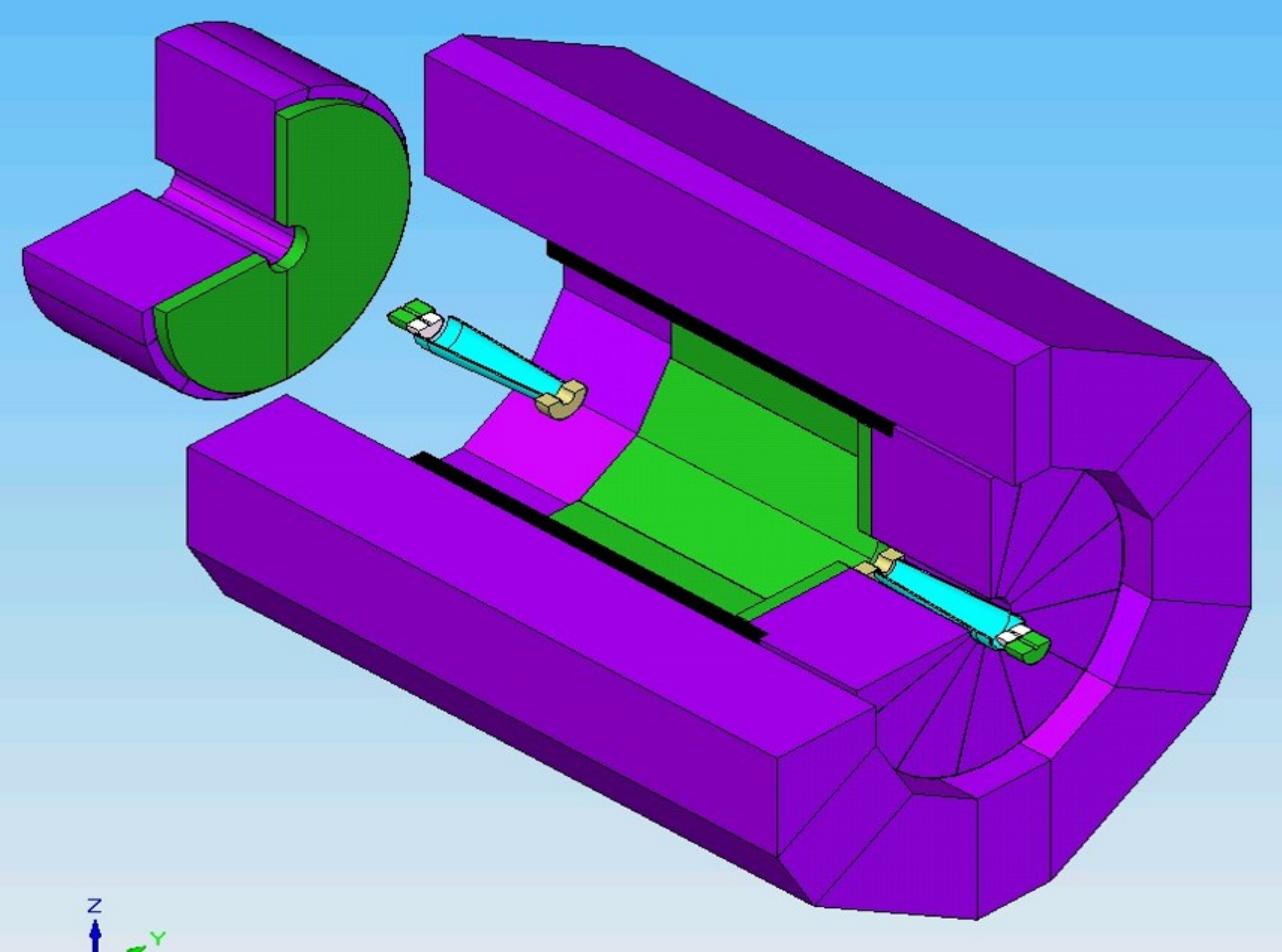}
\caption{Components of the SiD calorimeter system: purple - hadron calorimeter;
green - electromagnetic calorimeter.}
\label{fig:calorimeter}
\end{subfigure}%
\begin{subfigure}{.45\textwidth}
\centering
\includegraphics[width=.7\linewidth]{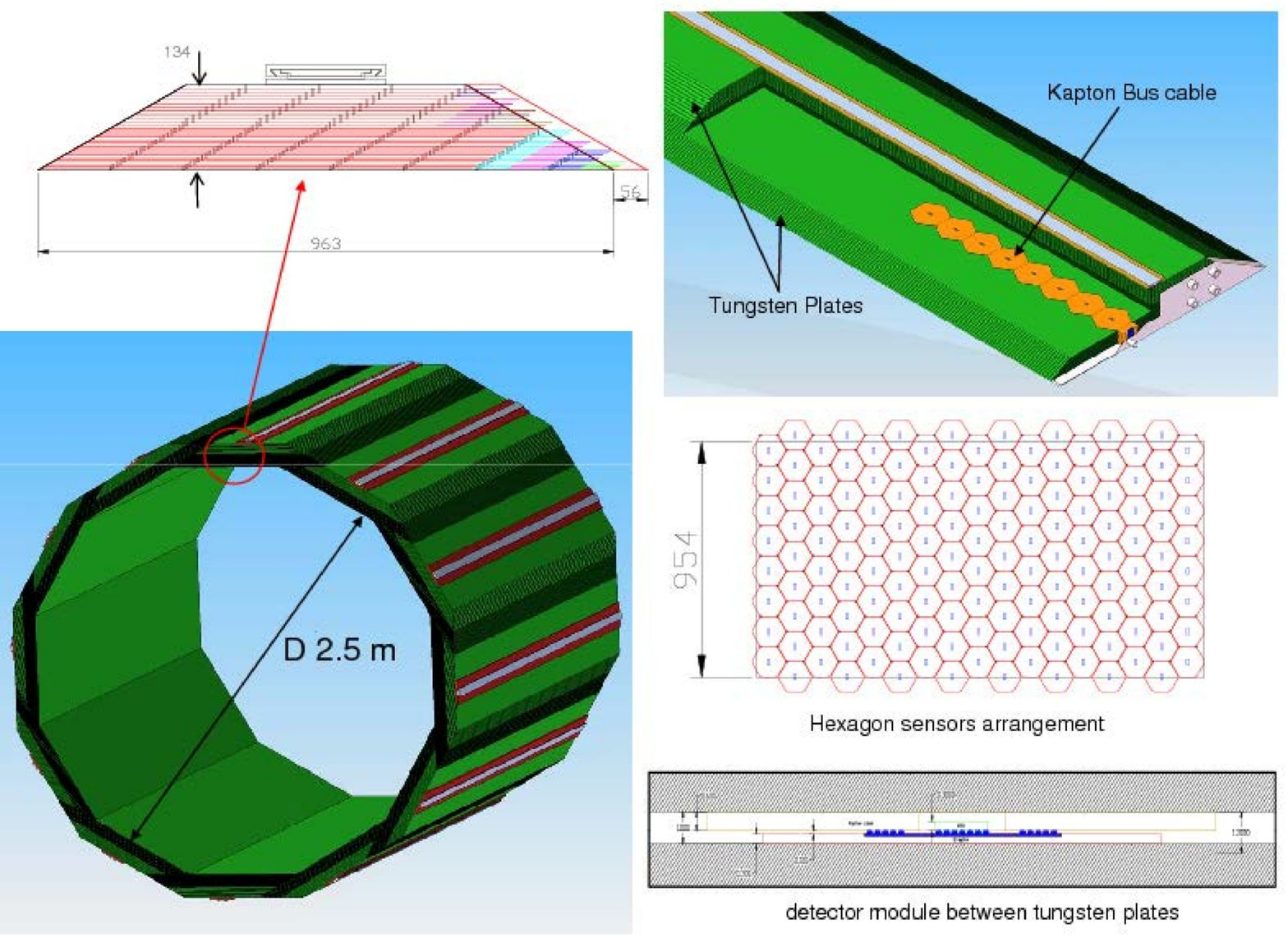}
\caption{Mechanics and components of the SiD electromagnetic calorimeter.}
\label{fig:ECAL}
\end{subfigure}
\caption{SiD Calorimeter system and detail of electromagnetic calorimeter.}
\label{fig:cal_ecal}
\end{figure}

SiD calorimetry is designed for the particle plow approach to improving jet energy
resolution. The goal is to achieve 3\% or better jet energy resolution for jets above 100 GeV.
For a PFA the tracker and calorimeter must work together to ensure efficient and effective
association of charged tracks with the correct energy deposits in the calorimeter. This
implies the need for a high degree of transverse and longitudinal segmentation in the 
calorimeters. The Moliere radius for the electromagnetic calorimeter should be minimized to
facilitate the separation of charged tracks and electromagnetic showers. Naturally, the entire 
calorimeter system is located inside the volume of the solenoid - which, however, imposes
limitations on radial dimensions due to cost considerations. 
The main elements of the calorimeter system are shown in Fig.~\ref{fig:cal_ecal} (a).

\subsection{Electromagnetic calorimeter}
In addition to supporting the PFA requirements described above, the ECAL must
allow precise measurement of electrons and positrons from Bhabha scattering for
determination of electroweak couplings and for a component of the measurement of
the luminosity spectrum. The ECAL should also provide for efficient detection 
and measurement of photons and pizero, for contributions to the jet energy resolution
and for reconstruction of $\tau$ decays. 
The ECAL baseline design \cite{ECal} features tungsten absorber plates and highly segmented silicon
sensor layers. The main elements of the design are shown in Fig.~\ref{fig:cal_ecal} (b).

Each silicon sensor is divided into 1024 pixels, read out by one KPiX ASIC \cite{KPiX}.
Each ECAL layer features an aggressive design with only 1.25 mm gap between 
successive tungsten plates - including the sensor, KPiX, flex cables and a 
passive cooling system. A nine-layer prototype of the ECAL has been tested and 
single and multiple electron tracks successfully recorded \cite{ECal}.
There also is a MAPS (Monolithic Active Pixel) design for the ECAL, which has
very fine, 50$\times$50 micronsquare, silicon pixels, for which first generation
sensors have been tested.

\subsection{Hadron calorimeter}
The HCAL is designed for efficient and unambiguous identification of energy
deposits by charged particles, their association with the related tracks in
the tracking system, and the measurement of the energies of neutral particles.
Within the PFA approach, these functions demand fine transverse and longitudinal
segmentation, with the requirement (for the barrel sections) to keep the active 
layer thickness to a minimum to control the cost of the radially external solenoid.

\begin{figure}
\centering
\begin{subfigure}{.45\textwidth}
\centering
\includegraphics[width=.7\linewidth]{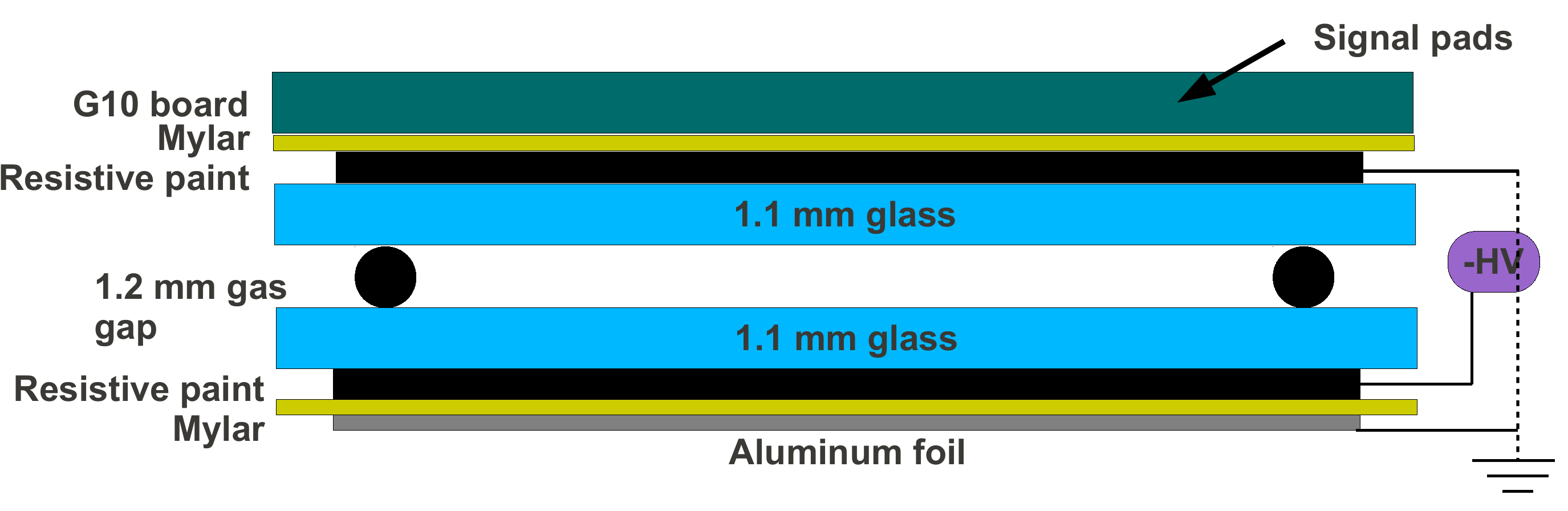}
\caption{Design of the two-glass RPC for the SiD hadron calorimeter.}
\label{fig:RPC_HCAL}
\end{subfigure}%
\begin{subfigure}{.45\textwidth}
\centering
\includegraphics[width=.7\linewidth]{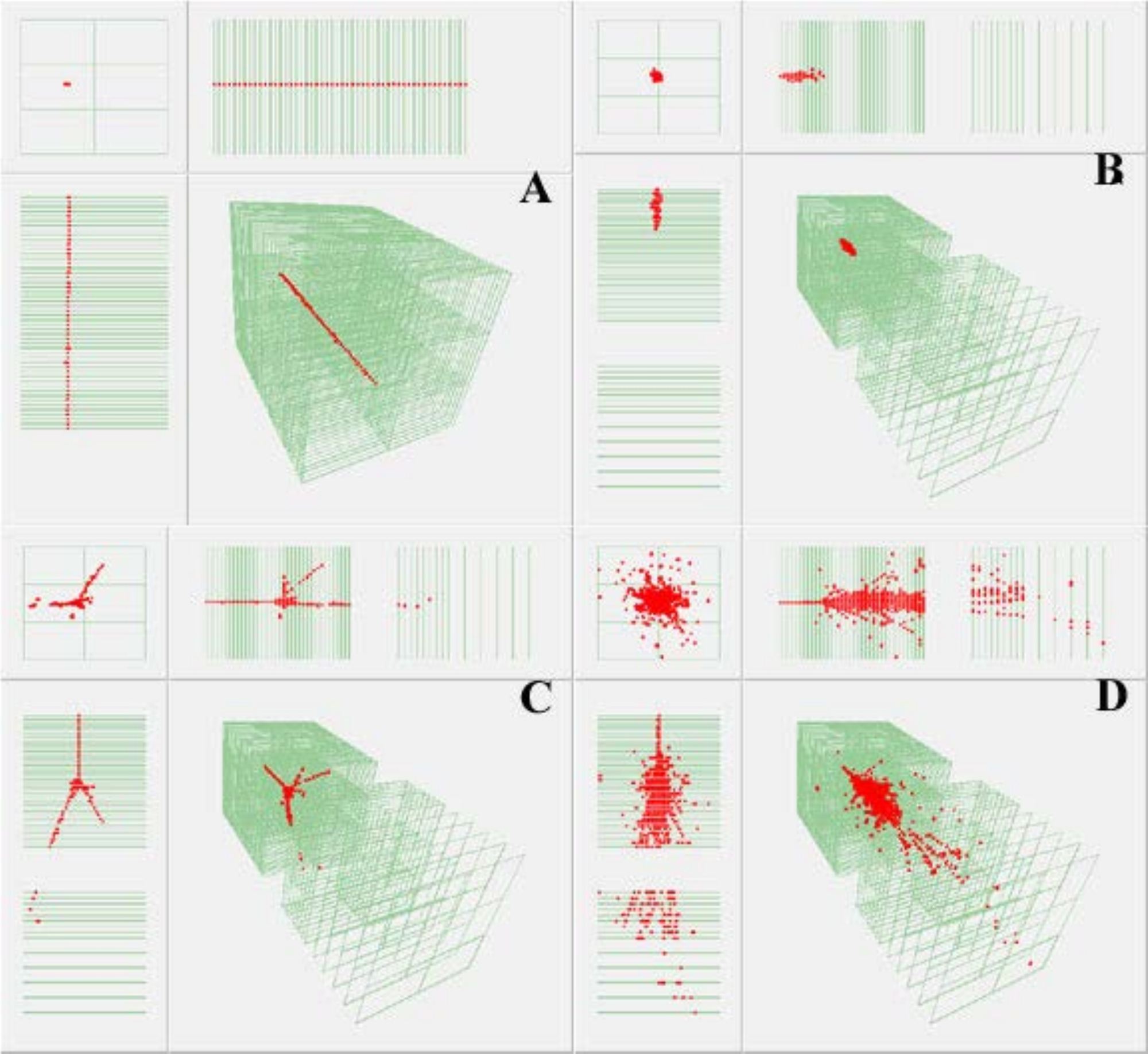}
\caption{Events recorded in the digital hadron calorimeter prototype.}
\label{fig:RPC_proto}
\end{subfigure}
\caption{The RPC design and prototype results for the SiD Hadronic Calorimeter.}
\label{fig:RPC_2}
\end{figure}

The baseline technology for the HCAL is resistive plate chambers (RPC) with steel
absorber plates \cite{RPC} . The basic RPC design is shown in Fig.~\ref{fig:RPC_2} (a).

\begin{figure}[htb]
\centering
\includegraphics[height=3.0in]{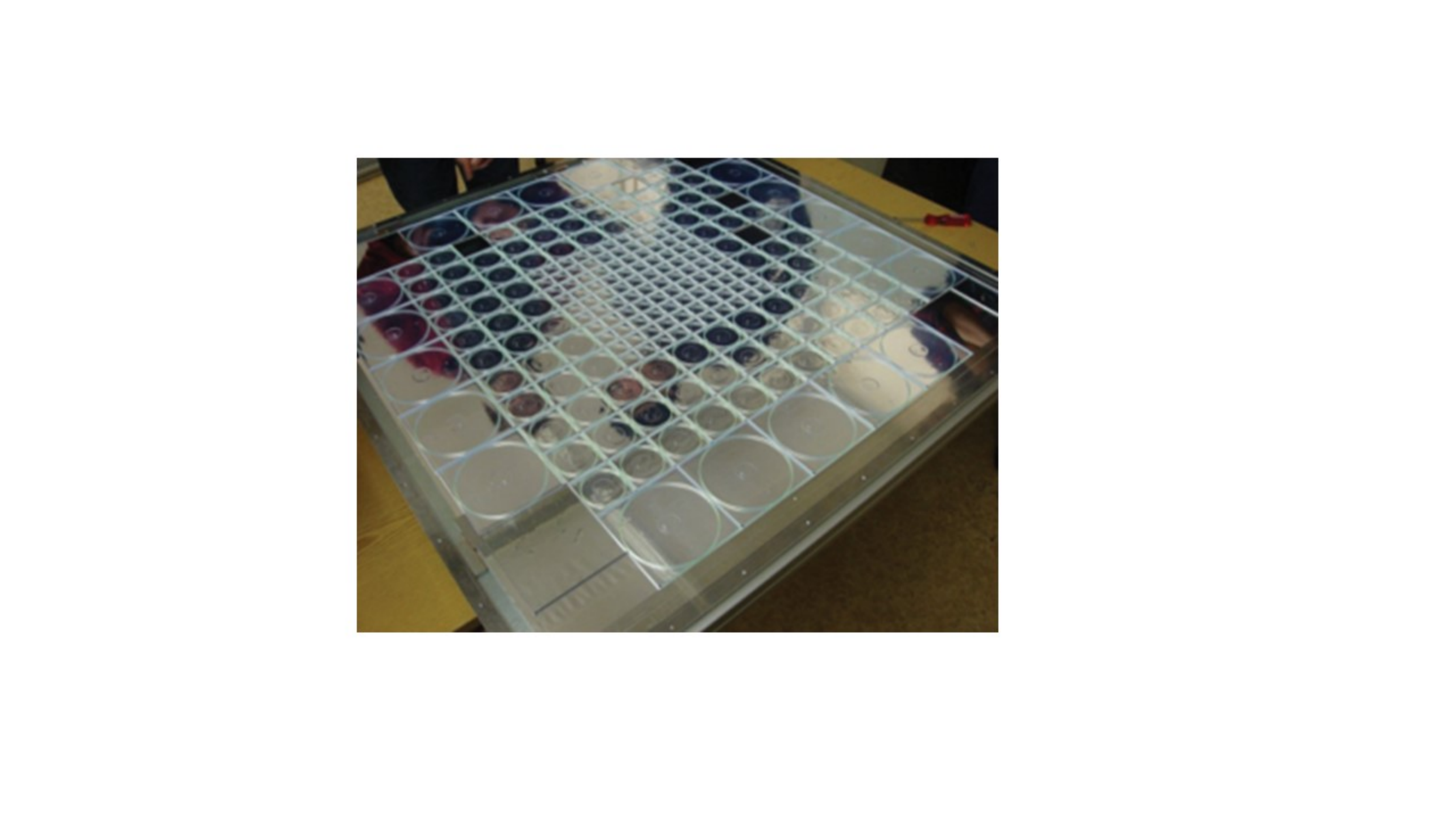}
\caption{Active layer from the scintillating tile-SiPM approach to the SiD HCAL.}
\label{fig:SCINT_HCAL}
\end{figure}

The baseline design has been implemented in a 38-layer prototype with transverse size 
1$\times$1~m$^2$, large enough to contain hadronic showers. Fig.~\ref{fig:RPC_2} (b)
shows several examples of hadron showers, and a single muon track, recorded in the prototype which had 
1 $\times$ 1~cm$^2$ readout pads. 
Several other technologies are also being developed and considered for the SiD HCAL: Scintillator tiles, 
GEM's (both foils and ThickGem), Micromegas, and variations on the RPC approach. Fig.~\ref{fig:SCINT_HCAL} 
shows an example: a 1$\times$1~m$^2$ active layer of scintillator tiles with SiPM readout \cite{AHCAL}.

\subsection{Forward Calorimetry}
\begin{figure}[htb]
\centering
\includegraphics[height=3.0in]{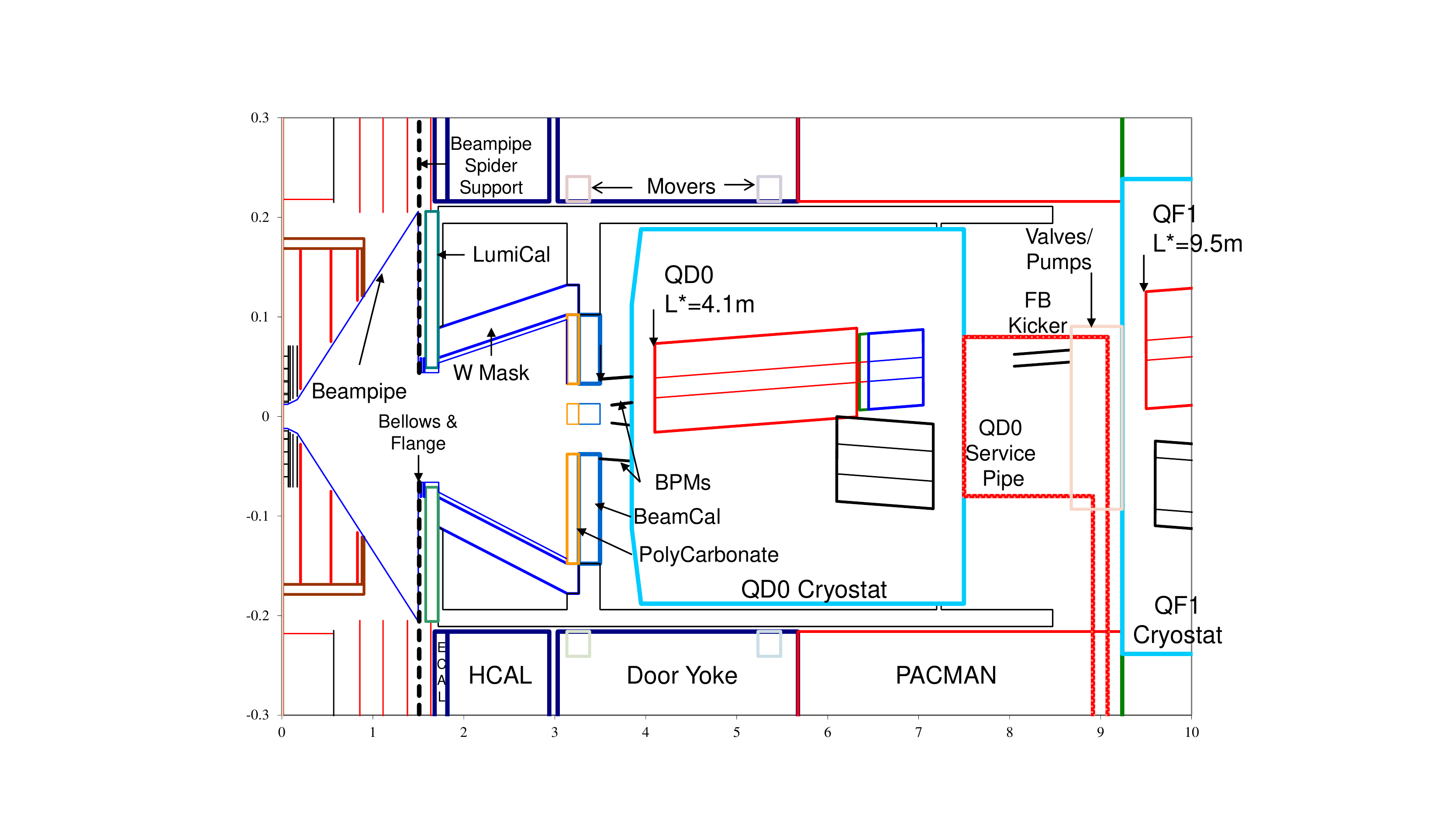}
\caption{The new forward region layout for SiD with L* of 4.1m.}
\label{fig:SiD_Forward}
\end{figure}

Fig.~\ref{fig:SiD_Forward} shows the recently updated forward calorimeter and 
luminosity calorimeter layout with the new common L* agreed by SiD and ILD.
The LumiCal will use small angle Bhabha scattering to determine the integrated
luminosity to better than one part per mil. The BeamCal will provide an instantaneous
measurement of the luminosity using beamstrahlung pairs, and will provide small-angle
coverage for physics searches.

\section{Muon System and Flux Return}
\begin{figure}
\centering
\begin{subfigure}{.45\textwidth}
\centering
\includegraphics[width=.7\linewidth]{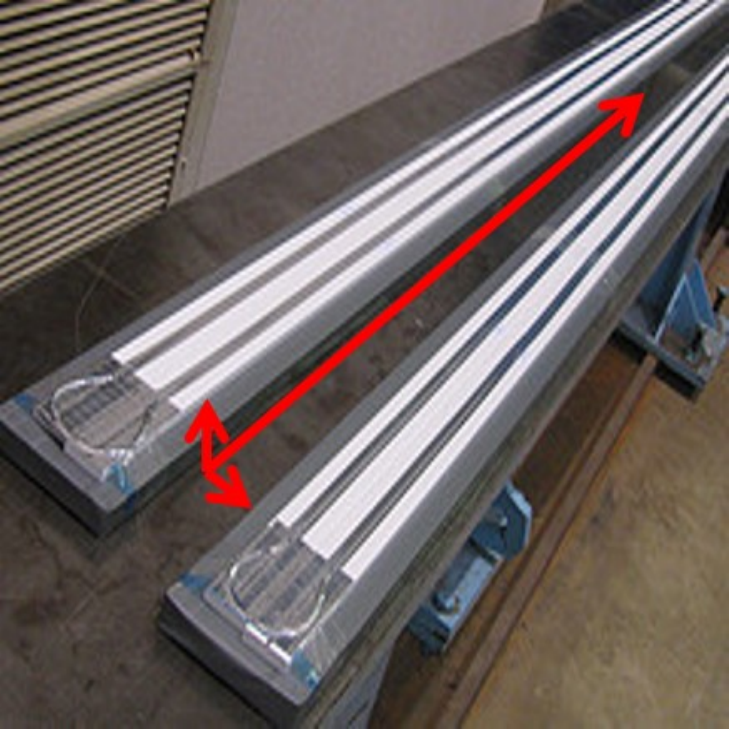}
\caption{Muon system prototype with long scintillator strips with embedded wavelength shifting fibers.}
\label{fig:SiD_Muon}
\end{subfigure}%
\begin{subfigure}{.75\textwidth}
\centering
\includegraphics[width=.7\linewidth]{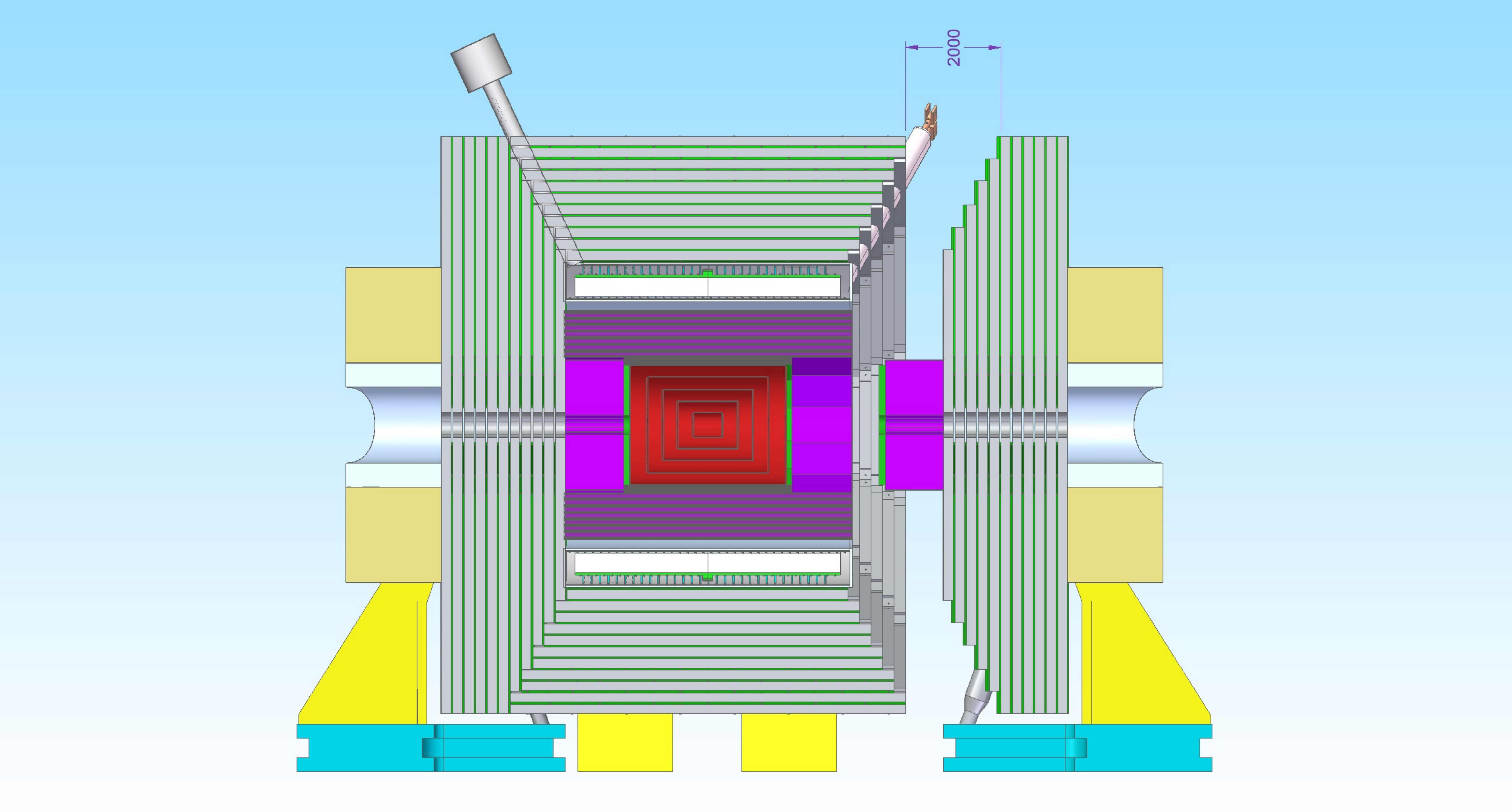}
\caption{New design of the SiD flux return steel.}
\label{fig:ECAL}
\end{subfigure}
\caption{Muon system steel and flux return, and scintillator prototype.}
\label{fig:SiD_Mu_Steel}
\end{figure}

The baseline technology for the muon system is long scintillator strips with wavelength
shifting fibers and SiPM readout. The roles of the muon system are to identify muons from
the interaction point efficiently, and, as a tail-catcher, to flag possible shower leakage through the
superconducting solenoid as a part of the particle flow algorithm input. Fig.~\ref{fig:SiD_Mu_Steel} (a)
shows prototype long strips and fibers underdevelopment.
The steel of the muon system acts as the flux return for the magnetic field from the superconducting 
solenoid. There is a local site requirement of a maximum of 50 Gauss at 15m from the main detector axis. 
Fig.~\ref{fig:SiD_Mu_Steel} (b) shows a recent design of the muon steel, with a 30 degree angle between the
barrel and endcap steel. This design limits the fringe field and satisfies the 50 Gauss requirement.

\section{Installation}
\begin{figure}[htb]
\centering
\includegraphics[height=2.0in]{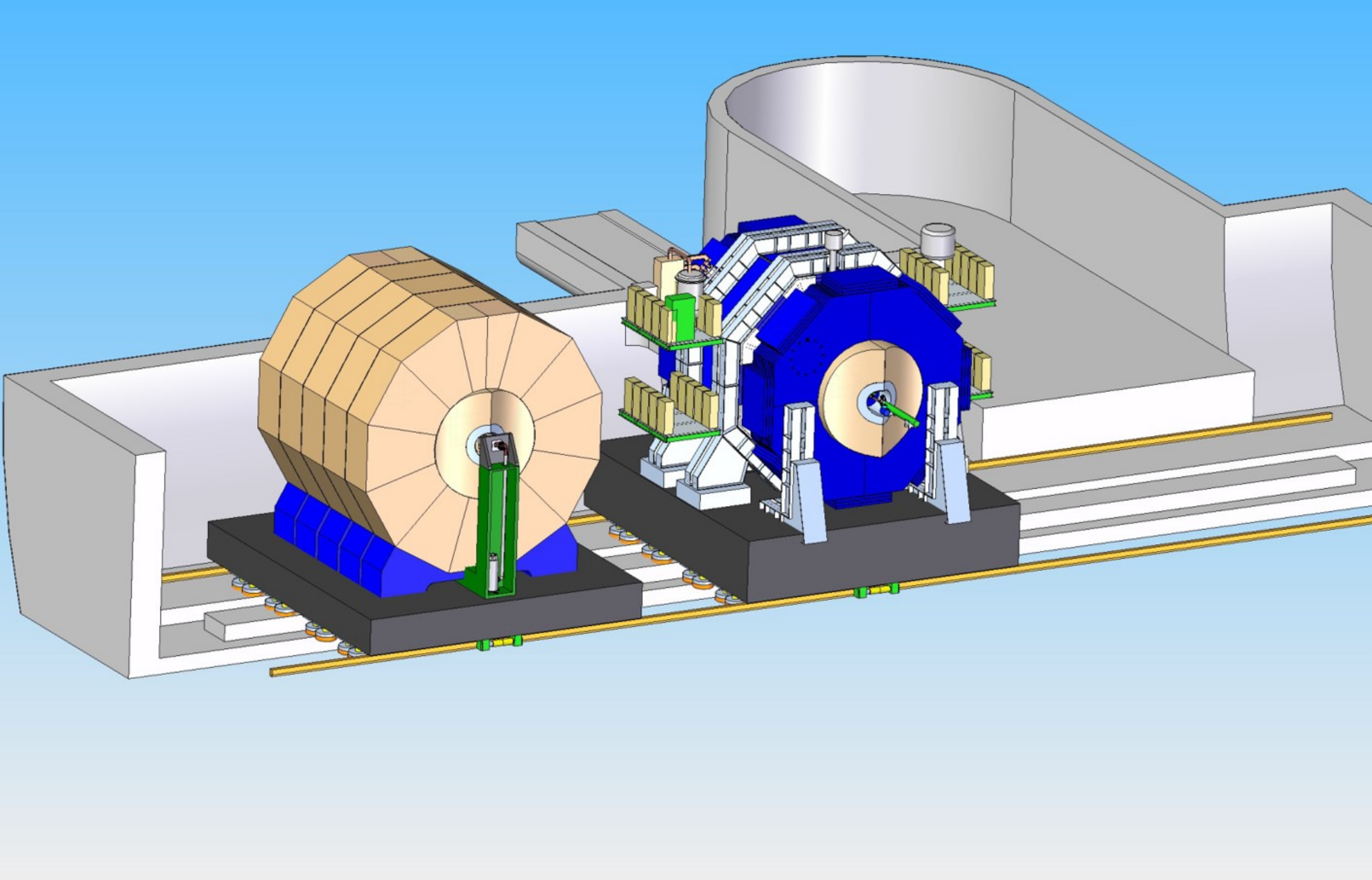}
\caption{The SiD Detector, on the beam axis, and the ILD Detector in the push-pull configuration.
The more compact design of the SiD Detector results in a deeper platform as shown in black.}
\label{fig:SiD_Push_Pull}
\end{figure}

The single interaction region of the ILC requires a sharing between the two detector concepts 
in a push-pull arrangement as shown in Fig.~\ref{fig:SiD_Push_Pull}. An assembly and installation
procedure has been created. It is estimated to take a period of eight years to complete.

\section{Summary}
The design of the SiD Detector Concept has been presented, with details of subsystems.
Physics studies with this design have shown that superb performance is expected on the
full range of ILC physics. Current detector development topics include a third generation 
Chronopix and advanced 3-D for the vertex detector, a new silicon sensor for the ECAL, engineering 
studies for a full-size scintillator-steel HCAL module, and overall optimization of the detector
design. The SiD Consortium remains open to new colleagues and to creative input to further
optimize or improve the detector design.



\begin{thebibliography}{99}


\bibitem{Chrono}
N. Sinev, talk at SiD Workshop, SLAC, January 2015.
https://agenda.linearcollider.org/event/6522/session/8/contribution/9
/material/slides/1.pdf

\bibitem{VIP_chip}
R.Lipton, talk at SiD Workshop, SLAC, January 2015.
https://agenda.linearcollider.org/event/6522/session/8/contribution/12
/material/slides/1.pdf

\bibitem{ECal}
R.Frey at SiD Workshop, SLAC, January 2015.
https://agenda.linearcollider.org/event/6522/session/8/contribution/15
/material/slides/0.pdf

\bibitem{KPiX}
D. Freytag et al., Linear Collider Power Distribution and Pulsing Workshop, Paris, 2011.
https://agenda.linearcollider.org/event/5010/session/6/contribution/14
/material/slides/1.pdf

\bibitem{RPC}
J.Repond, talk at LCWS 2013, U. Tokyo, November 2013.
http://agenda.linearcollider.org/event/6000/session/38/contribution/136
/material/slides/1.pdf

\bibitem{AHCAL}
B.Bilki et al.,  JINST 10 (2015) P04014.




\end{thebibliography}
\end{document}